\documentclass[twocolumn,showpacs,preprintnumbers,amsmath,amssymb]{revtex4}
\usepackage{tabularx,graphicx}\begin{document}
\newcommand{\beq}{\begin{equation}}
\newcommand{\eeq}{\end{equation}}
\newcommand{\beqn}{\begin{eqnarray}}
\newcommand{\eeqn}{\end{eqnarray}}
\newcommand{\bmath}{\begin{subequations}}
\newcommand{\emath}{\end{subequations}}

\title{ Electron-hole asymmetry and superconductivity}
\author{J. E. Hirsch }
\address{Department of Physics, University of California, San Diego\\
La Jolla, CA 92093-0319}

\date{\today} 

\begin{abstract} 
In a solid, transport of electricity can occur via electrons or via holes. In the normal state
no experiment can determine unambiguously whether the elementary mobile carriers have
positive or negative charge. This is no longer true in the superconducting state:
superconductors know the difference between electrons and holes. This indicates that
electron-hole asymmetry plays a fundamental role in superconductivity, as proposed
by the theory of hole superconductivity.
\end{abstract}
\maketitle 
 
\section{Introduction}
Imagine a parallel universe  where the heavy proton is negatively charged and the light 
electron  is positively charged. Solids in such a world will look the 
same as in ours. Solid state experimentalists would measure a positive Hall coefficient for 
Li and Au, and a negative Hall coefficient for Pb and Nb, a positive thermoelectric power
for Na and a negative one for In. Imagine furthermore that in this world
 they are unable to do experiments with
cathode ray tubes where the mobile charge carriers would escape from the solid and
propagate in free space. How would experimentalists in this world  be able to tell whether
the elementary charge carriers that are actually moving when electricity flows through a metal have positive
or negative charge?

The answer is, they would not if they can only do experiments in the normal state of the
metals and the carriers never leave the solid. There is no experiment in the normal
state that can determine unambiguously that the actual elementary particles that are moving in
the metal in this parallel universe have positive charge, and that the heavy
ions that are only rattling around their equilibrium positions have negative charge. All measurements would be equally
consistent with the assumption that in some materials (Pb, Nb, In) the transport is dominated by  elementary
mobile negative carriers and in others (Li, Au, Na) by the positive 'absence' of that carrier.

A hint that this assumption would be incorrect would be provided by the fact that 
the best conductors of electricity in this world, eg copper, have a positive Hall
coefficient, while the metals with negative Hall coefficient, eg Pb or Nb have a substantially
higher resistance. This  could be argued to suggest that the positive carriers are more
mobile than the negative ones, hence that the elementary charge carrier is more
likely to have a positive rather than a negative charge. However, theorists in this world would
counter-argue that such reasoning is not well founded
because the electronic band structure of materials is very complicated and the sign of
the Hall coefficient does not have a simple relationship with the sign of the
charge carriers.

Now in this world, as in ours, many materials become superconducting at low
temperatures. Somebody may observe that the vast majority of
superconducting materials have a negative Hall coefficient in the normal 
state (eg Pb, Nb, In, V, $MgB_2$,  $YBa_2Cu_3O_7$),
and that the best conductors with positive Hall coefficient never become
superconducting (eg Cu, Ag, Na, Ca). Furthermore, that the superconductors with the highest critical temperatures
are usually very bad conductors in the normal state. This could be taken to suggest  that the negative carriers
in the normal state of these superconducting materials have the 'wrong' sign, and that by going
superconducting the material somehow manages to avoid this situation and instead have the real mobile positive carriers do the
conduction    as in the 'good' normal state conductors Cu or Na. This hypothesis would be supported by the observation that   as the temperature is raised and approaches $T_c$ many
superconductors do show a positive Hall coefficient, that becomes negative again
as $T$ is raised above $T_c$\cite{sign1,sign2}. However, theorists in this world would argue against this that
first, the sign of the Hall coefficient has little to do with the sign of the charge carriers,
and second, that the sign reversal of the Hall coefficient in the mixed
state has to do with the complicated dynamics of vortex motion rather than with the
sign of the charge carriers\cite{twoband}.

In summary, without cathode ray tubes or the photoelectric effect or thermoionic emission or STM's or any
other experiment where the charge carrier is extracted from the solid it is
impossible to determine $unambiguously$ the sign of the mobile charge carriers in solids from
experiments in the normal state, as well as from many experiments in the superconducting state.
However, the observations discussed above  $suggest$, even if they don't $prove$,  that in this parallel world the sign of the elementary charge carrier 
in metals is positive, as they suggest that in our world it is negative. 

\section{The experiments}

Remarkably, there are experiments in the superconducting state that can determine the sign of the elementary
charge carriers in solids, as well as their mass, without the charge carriers ever
leaving the solid.

\subsection{The magnetic field of a rotating superconductor}

Consider a cylindrical superconductor with its axis along the $z$ direction. If this
superconductor is rotated with angular velocity $\vec{\omega}$ around its symmetry axis
a magnetic field develops in its interior, given by\cite{london0}
\beq
\vec{B}=-\frac{2mc}{e}\vec{\omega}
\eeq
where $e$ is the charge of the mobile charge carrier, with its sign, $m$ its mass and $c$ the speed of light. For
our world, where $e<0$, it means that $\vec{B}$ is in the same direction as $\vec{\omega}$.
In the parallel universe, $\vec{B}$ would point in direction opposite to $\vec{\omega}$.

\begin{figure}
\resizebox{6.5cm}{!}{\includegraphics[width=7cm]{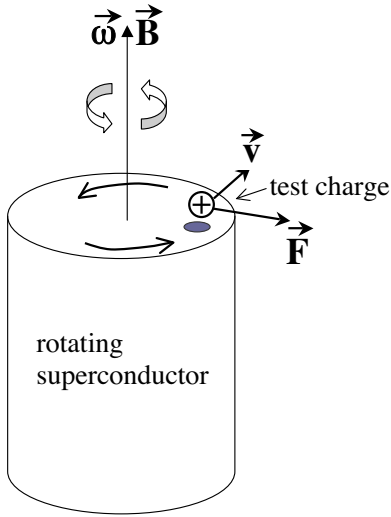}}
\caption{A superconductor rotating counterclockwise as seen from the top gives rise to a magnetic field pointing up,
parallel to $\vec{\omega}$, if the mobile carriers in the superconductor have $negative$ charge.
A $positive$ test charge in the figure moving with velocity $\vec{v}$ over the superconductor experiences an $outward$ force. }
\label{london1}
\end{figure}

Just in case in the parallel world they may use a different sign convention to define magnetic fields, let' s clarify
operationally what this means:  a test charge $q$ moving above the top surface of the
rotating superconductor in the same direction as the surface right below it 
will feel a Lorenz force that points $in$ if $q$ has $the$ $same$ sign as the mobile charge
carriers in the superconductor, and points $out$ if $q$ has $opposite$ sign as the mobile
charge carriers in the superconductor. In Figure 1, the force on the test charge is pointing out,
so the test charge is positive if Figure 1 depicts our world, and  negative if it depicts the
parallel world. This experiment determines unambiguously
the sign of the mobile charge carriers in the solid.

The magnetic field of a rotating superconductor has been measured for both
conventional (Pb \cite{pb}, Nb \cite{nb}, Sn \cite{sn} and Hg \cite{sn}),
semiconventional ($BaPb_{1-x}Bi_xO_3$ \cite{bapb}) and high $T_c$ ($YBa_2Cu_3O_7$ \cite{bapb}) superconductors. The relation Eq. (1) is found to hold,
with the sign discussed above and $m$ the $bare$ electron mass. The reason the
magnetic field points in the direction discussed is understood qualitatively as follows:
when the solid starts rotating the ions move and the superfluid electrons
'lag behind' in the region within a penetration depth of the surface; the result is a magnetic field pointing in the
direction given by the sign of the charge of the rotating ions, i.e. parallel to the angular
velocity of the solid in our world.

\begin{figure}
\resizebox{6.5cm}{!}{\includegraphics[width=7cm]{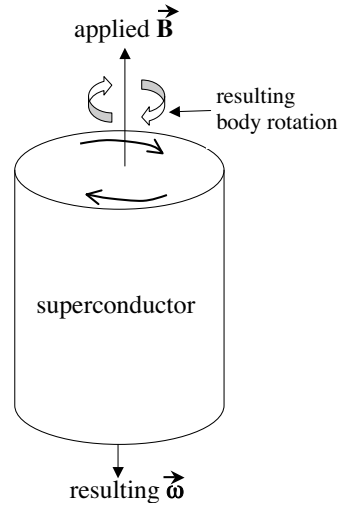}}
\caption{When a magnetic field is suddenly turned on pointing $up$ the superconductor will start to rotate $clockwise$ as seen
from the top, i.e. with angular velocity $antiparallel$ to the applied field, if the mobile carriers have $negative$ charge.}
\label{london1}
\end{figure}

\subsection{The gyromagnetic effect}

A related effect results if a magnetic field is suddenly applied to a superconductor.
Consider again a cylindrical superconductor with axis along the $z$ direction, at rest initially, and
apply a magnetic field in the positive $z$ direction. The supercurrent that develops 
to nullify the magnetic field in the interior (Meissner effect) 
will have the negative electrons rotating with angular momentum pointing in the $+z$
direction. For the total angular momentum of the cylinder to be unchanged,
the cylinder will start rotating with angular momentum in the $-z$ direction, as shown in Figure 2. Once again the result 
would be opposite if the elementary charge carriers giving rise to the supercurrent had positive charge.
This experiment has been done in conventional superconductors\cite{gyro1,gyro2} and not, to our
knowledge, in high $T_c$ superconductors. There is little doubt however that the
result would be the same. 

In summary, superconductors know for a fact that the elementary charges that carry the supercurrent are
negative in our world. Normal metals instead exhibit properties consistent with both negative and positive charges
being the charge carriers.

\section{Superconductivity and electron-hole asymmetry}

The existence of the (conventionally called) London field Eq. (1) in the interior of a rotating superconductor can be derived
from the London equation
\beq
\vec{\nabla}\times\vec{v}_s=-\frac{e_s}{m_sc}\vec{B}
\eeq
where $e_s$, $m_s$ and $\vec{v}_s$ are the charge, mass and velocity of the superfluid carries, assuming that in the interior of the
superconductor the superfluid is rotating at the same velocity as the lattice,
\beq
\vec{v}_s=\vec{\omega}\times\vec{r}
\eeq
hence $\vec{\nabla}\times\vec{v}_s=2\vec{\omega}$. The London penetration depth
\beq
\lambda=(\frac{m_s c^2}{4\pi n_s e_s^2})^{1/2}
\eeq
is obtained from Eq. (2) using the expression for the supercurrent $\vec{j}_s=n_s e_s \vec{v_s}$ together with Ampere's law
$\vec{\nabla}\times\vec{B}=(4\pi/c) \vec{j_s}$.

Conventional descriptions of superconductivity use electron-hole symmetric models, hence they cannot predict an effect
as profoundly electron-hole asymmetric as the London field. In a model superconductor with hole carriers
in the normal state, eg an attractive Hubbard model with the band almost full, in the London penetration
depth Eq. (4) $n_s$, $e_s$ and $m_s$ refer to the density, charge and mass of superfluid $hole$ carriers. Of course Eq. (4) is independent
of the sign of $e_s$, but Eq. (2) from where it was derived is not.

So where is electron-hole symmetry broken in the London equations? Eq. (2) would still be valid in a hole description in terms of
\bmath
\beq
\vec{v}_{s,hole}=-\vec{v}_{s,electron}
\eeq
\beq
e_{s,hole}=-e_{s,electron}
\eeq
\emath
however Eq. (3) would $not$ hold for $\vec{v}_{s,hole}$. This is because Eq. (5a) is valid for the $relative$ velocity of electrons and holes with
respect to the crystal structure; instead, in writing Eq. (3) one is stating that $\vec{v}_s$ is the $absolute$ superfluid
velocity, independent of the velocity of the crystal structure. For the absolute velocity of electrons and holes, Eq. (5a) does not hold when the crystal is moving and hence
Eq. (2) for holes does not hold .

It is also important to note that the mass entering the London field Eq. (1) is experimentally determined to be the $bare$ electron mass. Instead
in the expression for the
London penetration depth Eq. (4), $m_s$ is conventionally interpreted to be the effective mass of the hole carrier in the
normal state, which is usually larger than the bare electron mass due to 'dressing' effects from electron-electron and
electron-lattice interactions.

So the puzzle is then: how do  dressed hole carriers in the normal state with a positive charge and a large effective mass become
undressed electron carriers in the superconducting state with a small (bare) mass? An answer is provided by the theory
of hole superconductivity.

\section{Relation with the theory of hole superconductivity}

\begin{figure}
\resizebox{8cm}{!}{\includegraphics[width=9cm]{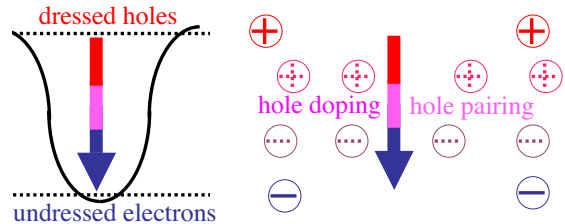}}
\caption{Phenomenology of hole superconductivity. As the Fermi level goes up
in the band, bare electrons become dressed holes. When holes pair it is as if locally 
the band becomes less full, hence holes undress and turn into electrons.}
\label{london3}
\end{figure}

The theory of hole superconductivity proposes that superconductivity originates in the
fundamental asymmetry between electrons and holes in condensed matter\cite{hole}.
It points out that in electronic energy bands carriers are undressed and light when the
Fermi level is close to the bottom of the band (electrons) and that they are
highly dressed and heavy when the Fermi level is close to the top of
the band (holes)\cite{hole1,hole2}, as shown schematically  in Figure 3. Furthermore that carriers have low kinetic energy when they are at
the bottom of the band (electrons) and high kinetic energy when they are at
the top of the band (holes)\cite{hole3}. The energetics that drives superconductivity is kinetic energy lowering,
or equivalently effective mass reduction, of the hole carriers when they pair\cite{hole4}.
When holes pair, the local density of holes increases, hence the
band becomes locally less full and the carriers become  more electron-like and less dressed\cite{hole10}.
In calculating the London penetration depth in the theory\cite{hole5}, it is found that the effective mass that enters is
$smaller$ than the large effective mass of the hole carriers in the normal state, hence the London penetration depth is
smaller than expected from the normal state effective mass. This in turn
gives rise to low frequency optical sum rule violation\cite{hole6} and to color change\cite{hole7}: as the effective mass in the superconducting state is smaller
optical spectral weight is transfered down from high frequencies to the $\delta-$function that determines the
London penetration depth. These effects have recently been seen experimentally\cite{santander,marel}.
The fact that the carriers that carry the supercurrent near the surface are electrons rather than
holes implied by Eq. (1) is also predicted by the theory\cite{hole8}.

Thus the theory of hole superconductivity in its present formulation conservatively asserts that dressed hole carriers
partially undress, their effective mass decreases and they partially turn into electrons when they become 
superconducting. Nature is bolder than that: experiments that measure Eq. (1) tell us that this happens all 
the way, the dressed hole carriers become
undressed free electrons, with the free electron mass and charge, in the superfluid
state. 

\section{Discussion}

In contemporary solid state physics there is a common practice, especially
among many-body theorists, to regard electrons and holes as equivalent quasiparticles and hence to use 
electron-hole symmetric Hamiltonians to describe physical systems. When a particular model or a particular situation
does not accommodate that expectation it is  often  stated that such model or situation 'violates' electron-hole symmetry,
as if that would be a bad or   unphysical or at least an unusual  thing to do. The experiments and observations
discussed above however underline the fact that there is 
a profound asymmetry between electrons and holes in condensed matter. The facts that superconductors in the normal
state exhibit almost always dominantly hole carrier transport\cite{born,feynman,chapnik}, and that electron-hole  asymmetry shows up clearly
and unambiguously in the superconducting state,   suggest that there is an intimate relationship between electron-hole asymmetry
and superconductivity. The theory of hole superconductivity rests on the proposition that electron-hole asymmetry is the key to superconductivity\cite{hole9}.

\acknowledgements
The author is grateful to the Instituto de Ciencia de Materiales de Madrid, CSIC, for its hospitality while this
work was performed.

 \end{document}